\documentclass{article}

\usepackage{arxiv}

\usepackage[utf8]{inputenc}
\usepackage[T1]{fontenc}
\usepackage{hyperref}
\usepackage{url}
\usepackage{booktabs}
\usepackage{amsmath,amssymb,amsfonts,amsthm}
\usepackage{nicefrac}
\usepackage{microtype}
\usepackage{graphicx}
\usepackage{xcolor}
\usepackage{bm}
\usepackage{bbm}
\usepackage{natbib}

\graphicspath{{Fig_submit/}}

\theoremstyle{plain}

\numberwithin{equation}{section}

\newcommand{\blue}[1]{{\color{black}{#1}}}

\title{Mathematical modeling to study combination strategies for neural stem cell therapy in glioblastoma}

\author{
Ashley Harrison \\
School of Mathematics and Statistical Sciences \\
Arizona State University \\
Tempe, AZ 85281, U.S.A.
\And
Austin Hansen, Anthony E. Bishay \\
Department of Mathematics \\
University of California, Riverside \\
Riverside, CA 92521, U.S.A.
\And
Greg Grigoryan, Vikram Adhikarla, Russell C. Rockne, Margarita Gutova  \\
Departments of Stem Cell Biology and Regenerative Medicine, \\
and Computational and Quantitative Medicine \\
Beckman Research Institute, City of Hope \\
Duarte, CA 91010, U.S.A.
\And
Heyrim Cho \\
School of Mathematics and Statistical Sciences \\
Arizona State University \\
Tempe, AZ 85281, U.S.A. \\
\texttt{heyrim.cho@asu.edu}
}

\begin{document}
\maketitle

\begin{abstract}
Neural stem cells (NSCs) are a promising treatment platform due to their ability to cross the restrictive blood-brain barrier and migrate toward damaged regions of the brain. These properties can be leveraged to target brain tumors, as NSCs can infiltrate tumor lesions and deliver therapeutic agents such as drugs, oncolytic viruses, and genetic materials. Despite the advantages of NSCs, translational challenges remain, including patient-specific chemotactic signaling and immune responses, variability in migration and localization at tumor sites, and optimization of delivery strategies. In this work, we focus on intranasal delivery of NSCs and combination strategies for the treatment of brain cancer. To investigate strategies for improving NSC arrival at the tumor site, we developed a hybrid mathematical model of NSC migration that couples an agent-based model with a partial differential equation, where cell motion is guided by the spatial dynamics of chemoattractants. Using data collected from a murine model and our mathematical model, we \blue{investigate treatment-motivated downstream effects associated with three combination strategies:} (1) IL-10 pretreatment, (2) CXCL12 pretreatment, and (3) radiotherapy. We find that IL-10 pretreatment improves NSC arrival numbers at tumor only under the Gaussian chemoattractant profile, whereas CXCL12 pretreatment moderately improves arrival numbers when the tumor is located in the anterior brain. Radiotherapy significantly enhances the tumor arrival under uniform profiles. These simulation studies suggest that both treatment strategies and chemoattractant distribution influence intranasal NSC delivery, and the proposed model can be used to predict how targeted modulation of chemotactic signaling and tissue-guided migration may improve delivery outcomes.
\end{abstract}


\section{Introduction} \label{sec:intro}


Neural Stem Cells (NSCs) have a number of unique capabilities that can be leveraged in treating a variety of different neurological pathologies, such as brain tumor, traumatic brain injuries, Alzheimer's, Parkinson's, Huntington's, and multiple sclerosis  \cite{rahimi_darehbagh_stem_2024,yang_therapeutic_2024,zahr2019translating,LiGutova2016,amirbekyan2023}. Among these is the ability to respond to chemical gradients in the brain, travel directly to tumor sites, and carry therapeutic agents \cite{barish2017,Gutova2019,Gutova2015}. 
%
Particular interest has been shown in the effects of NSCs on various glioblastoma, an aggressive form of cancer arising from glial cells in the central nervous system. While our understanding of the pathophysiology and etiology of glioblastoma has improved, effective treatments have stagnated in comparison. Prognosis for glioblastoma remains extremely poor, with a median survival of 15 months and a 5.5\% 5-year post-diagnosis survival rate. Thus, the importance of efficacious treatments has only increased in recent years, and NSC treatment seems to be a promising avenue.
To date, tremendous research has been undertaken to better understand the tumor-tropic properties of NSCs \cite{barish2017,Gutova2019,Gutova2021,zhao2008,Gutova2008,Gutova2015}. Due to their ability to cross the blood–brain barrier (BBB) and their strong migratory capacity and therapeutic potential, NSCs are considered effective vehicles for delivering therapeutic agents to the brain.

The tumor tropism of NSCs has been studied with multiple routes of delivery, including intracerebral and intravenous injection, where robust and reliable NSC homing to tumor site has been consistently observed \cite{barish2017,Gutova2019,Gutova2021,Gutova2015}. 
For example, Barish et al. \cite{barish2017} reported that 10–60\% of intracerebrally injected NSCs localized to the tumor, whereas intravenous injections yielded approximately 1.5\% arrival. Gutova et al. \cite{Gutova2015} studied tumor-directed migration of NSCs using intranasal delivery, and approximately 0.4–1.3\% of administered NSCs arrived at the tumor site within 1–4 days. Although intranasal injection achieves tumor arrival rates comparable to intravenous administration, it exhibits both lower overall efficiency and greater variability in the tumor site arrival numbers. 
Our previous computational study \cite{gomez_mathematical_2022} further supports this distinction that intranasal delivery yields fewer NSCs at the tumor compared to intracerebral injection, since intranasally administered cells must travel significantly longer anatomical pathways before encountering chemotactic cues, whereas intracerebral injection places NSCs directly on or near white-matter tracts that provide a fast migratory route to the tumor. 
These observations motivate the need to understand the migratory path of intranasally delivered NSCs and to predict the efficacy of NSC migration to the target site, as well as to develop combination strategies to improve therapeutic outcomes. \blue{Chemoattractants are soluble signaling molecules, including cytokines, particularly chemokines, as well as growth factors, that form spatial concentration gradients and direct cell migration toward regions of higher concentration through chemotaxis.} Therefore, this paper investigates combination treatment strategies aimed at modulating \blue{chemoattractants} dynamics to enhance intranasal NSC targeting efficiency, addressing insufficient cytokine-driven chemotactic signaling that may limit tumor tropism.

Prior modeling works in \cite{gomez_mathematical_2022} and \cite{rockne_long-term_2018} model NSC migration using structure-tensor coherence to distinguish white and gray matter regions in the brain, which captures tissue anisotropy. This resulted in a model that produced directed motion along principal white matter fibers and random movement in gray matter, with chemotactic bias when in a region above a set threshold for white matter. More recently, modeling was done \cite{hansen2025NSC} using advanced imaging techniques to resolve multiple crossing fiber directions, and represents NSC movement as a probabilistic combination of directed and random motion determined by local quantitative anisotropy. We extend the structure-tensor model by coupling the NSC migration agent-based model developed there with a chemoattractant field that evolves in time according to \blue{a time-dependent partial differential equation (PDE)}, which allows dynamic chemotactic cues to influence cell trajectories. 
%
\blue{
In section \ref{sec:method}, we present the hybrid model of NSC migration. 
In section \ref{sec:data}, we describe the experimental data and baseline simulations of NSC treatment without combination therapy. In section \ref{sec:result}, we investigate three combination treatment strategies. Finally in section \ref{sec:conclusion}, we summarize the main findings and discuss their implications. 
}

\section{Mathematical Model} \label{sec:method}

We present a mathematical model that captures the migration of therapeutic NSCs in the brain with cancer. The model integrates an agent-based framework to represent the dynamics of individual NSCs with \blue{an anisotropic} reaction-diffusion PDE that governs the evolution of the chemoattractant concentration driving directed migration. 
%

Let us first describe the agent-based model. 
It is known that cells, including tumor cells, show preferential migration along white matter tracts, which has led to recent models of glioblastoma tumor growth to incorporate the anisotropic nature of the brain \cite{wang2019invasion}. This can be achieved by utilizing diffusion tensor imaging (DTI), which captures the orientation of tissue fibers by measuring the anisotropic diffusivity of water within the brain. In addition, \cite{rockne_long-term_2018} has shown that therapeutic NSCs injected in a mouse brain cluster around white matter grey matter interfaces, and they found NSC clusters are often aligned with white matter tracts. Furthermore, a computational model that incorporated directed motion in white matter and random motion in gray matter as the primary migratory mechanisms showed good consistency with experimental results of NSC migration in naive mice \cite{hansen2025NSC}. Thus, we also assume that NSCs show preferential motion along white matter tracts, while exhibiting random motion in other regions. Fractional anisotropy ($FA$) is used in the model to distinguish between white and gray matter. It is an index determined from DTI data that quantifies the amount of anisotropy in white matter. It is computed in three-dimensions as
\begin{equation}
    FA = \sqrt{\frac{3}{2}} \frac{\sqrt{(\lambda_1-\hat{\lambda})^2 + (\lambda_2-\hat{\lambda})^2 + (\lambda_3-\hat{\lambda})^2}}{\sqrt{\lambda_1^2 + \lambda_2^2 + \lambda_3^2}}
\end{equation} where $\lambda_1$, $\lambda_2$, and $\lambda_3$ are the eigenvalues of the $3\times 3$ diffusion tensor, and $\hat{\lambda}$ is the average of those three eigenvalues. We define a point $\bm x = (x,y,z)$ to be white matter if $FA(\bm x)\geq \epsilon_{M}$ for some threshold value $\epsilon_{M}$, and grey matter otherwise. With this threshold, we can define the migration velocity of a NSC at location $\bm x^k$ as 
\begin{equation} 
\begin{cases} 
 \pm d_w \, \bm q( \bm x^k ) , \hspace{.5cm}& FA( \bm x^k ) \geq \epsilon_M, \qquad \textrm{on white matter} \\
 \,\, d_g \, \bm \xi(\omega) , \hspace{.5cm}& FA( \bm x^k ) < \epsilon_M, \qquad \textrm{on gray matter}
\end{cases}. 
\label{eq:model} 
\end{equation}
Here, the directed motion on the white matter is determined by $\bm q(\bm x)={e_1(\bm x)}/{\|e_1(\bm x)\|}$, which is the normalized eigenvector $e_1$ corresponding to the largest eigenvector of the diffusion tensor. In addition, $d_w$ and $d_g$ are the step sizes on the white matter and gray matter, respectively, and $\xi(\omega) \in \mathbb{R}^3$ is a random unit vector used to determine the direction of motion. In this paper, we represent random variables as functions of $\omega \in \Omega$ where $\Omega$ denotes the underlying probability space.

In addition to this, it is also shown that NSCs are able to respond to chemical gradients within the brain. In particular, the previous work in \cite{Gutova2008} has studied NSC migration in response to glioblastoma and has shown that uPA and other cytokines have a large effect on the migration of NSCs, with motion in the direction of increasing gradients. Thus, our agent-based model includes the effects of chemotaxis $\nabla C_j(t, \bm x(t))$ where $C_j(t, \bm x)$ is the $j$-th \blue{chemoattractant} concentration.

The \blue{stochastic differential equation} of the $ k$-th cell is written as  
\begin{equation} 
\blue{
 d X_t^k= \left[ \overbrace{ \psi_k  \, \bm b( X_t^k, S_t^k ) }^{\textrm{\,\,tissue-guided motility}} + \overbrace{ \eta_k  \sum_j \lambda_{c_j} \nabla C_j(t, X_t^k ) }^{ \textrm{chemotaxis} } \right] d t + \overbrace{ \psi_k \bm \Sigma(X_t^k) }^{\textrm{random motility}} d W_t,
 }
\label{eq:model1_}
\end{equation} 
\blue{where the white matter guided motility and random motility on the gray matter is defined as }
\begin{equation} 
\blue{\bm b( x, s ) = d_w \, s \, \bm q(x) 
\mathbbm{1}_{{\mathrm{FA}( x)\geq\varepsilon_M}}, 
\quad 
\bm \Sigma(x) = d_g \mathbbm{1}_{{\mathrm{FA}( x)< \varepsilon_M}}. 
}
\end{equation} 
There are two possible directions in the directed migration corresponding to the largest eigenvalue. We chose the direction \blue{$s \in \{ -1, 1\}$} that is consistent with the previous direction of motion by using the sign of the inner product between \blue{$d X_t^k / dt$ and $\bm q( X_t^k)$, in the discretized form, $S_{t_i}^{k} = \textrm{sign} [ ( x_i^k-x_{i-1}^k ) \cdot \bm q ( x_i^k ) ]$}.  
\blue{We define $\textrm{sign}[0]=1$, that is, when $ ( x_i^k-x_{i-1}^k ) \cdot \bm q ( x_i^k )=0$, the $+ \bm q (x_i)$ direction is selected.} 
The second term models the bias in NSC movement in the direction of increasing \blue{chemoattractant} concentration. $\lambda_{c_j}$ is the chemotactic sensitivity parameter that determines how strongly the \blue{chemoattractant} cues influence the direction and magnitude of NSC migration. 
Furthermore, stochasticity \blue{due to cell-to-cell variability} is incorporated into both tissue-guided and chemotactic migration through random variables  $\psi_k$ and $\eta_k$. We observe that while some cells migrate long distances, others remain near the injection site \cite{rockne_long-term_2018}. To capture this variability, we introduce stochasticity in both migration speed and chemotactic sensitivity through these parameters. A beta distribution was used to flexibly control the shape of the distribution on the bounded range $[0,\,1]$, allowing bias towards distinct motility regimes, as done in \cite{gomez_mathematical_2022}. In particular, $\blue{\psi_k(\omega) \sim \textrm{Beta}( 1,\, \beta_w ) }$ models tissue-guided motility, while $\blue{\eta_k(\omega) \sim \textrm{Beta}( \alpha_c ,\, 1 )}$ models chemosensitivity.

To describe the evolution of the chemoattractant, we consider the concentration of two chemoattractants, VEGF and MCP-1, to evolve with time as they diffuse, increase, and diminish due to their interaction with the NSCs. These two cytokines are well-established mediators of NSC migration. MCP-1 is an important chemotactic factor that directs the movement of neural stem and progenitor cells through CCR2 signaling and is frequently up-regulated in gliomas \cite{widera2004mcp}. VEGF, in addition to its angiogenic role, also acts as a chemoattractant for NSCs and is elevated in tumor regions, creating spatial gradients that can influence NSC trafficking \cite{SCHMIDT200924}. Because these cytokines shape the chemotactic landscape encountered by migrating NSCs, incorporating MCP-1 and VEGF enables the model to more accurately represent the biochemical cues governing NSC migration towards tumor. 
We consider a reaction-diffusion equation to describe the evolution of the chemoattractant $C_j(t,\bm x)$ as in \cite{mclennan2015} stated with modifications for the three-dimensional case and with Fokker-Planck diffusion operator for the anisotropic diffusion \cite{painter2013,swan2018patient}. In particular, we consider two chemoattractants, VEGF as $C_1(t,\bm x)$ and MCP-1 as $C_2(t,\bm x)$.  
\begin{equation}\label{PDE}
    \frac{\partial C_j}{\partial t} =\nabla \cdot \nabla \cdot ( \mathbf{D} \  C_j) - C_j \sum_{k=1}^{N(t)} \frac{\gamma}{(2\pi R^2)^{\frac{3}{2}}}\text{exp}\left(-\frac{\| \bm x - \bm x^k \|_2^2 }{2R^2}\right) + \blue{p(\bm{x}, C_j)}
\end{equation}
subject to the Neumann-zero boundary condition:
\begin{equation} \label{boundaryCond}
    \left.\frac{\partial C_j}{\partial \mathbf{n}} \right\vert_{(x,y,z)\in \partial \Omega_1} = 0
\end{equation}

Here we have defined $\Omega_1$ to be the domain of the three-dimensional mouse brain. 
$\gamma$ is the chemoattractant internalization rate, \blue{and} $R$ is the cell radius. 
In addition to these parameters, we define the diffusion tensor $\mathbf{D}$ as follows:
\begin{equation} \mathbf{D}(x) = 
    \begin{cases}
        D_w  \ v(x) \ v(x)^T \ \ \ \ \ \ \ \text{if} \ \ x\in \text{white matter} \\
        D_g  \ \mathbf{I} \ \ \ \ \ \ \ \text{if} \ \ x\in \text{gray matter}
    \end{cases}
\end{equation}
where $D_w$ and $D_g$ are the diffusion coefficients in white and gray matter, respectively. Diffusion tensor imaging yields, for every voxel, the eigenvectors characterizing the dominant diffusion direction. We normalize the principal eigenvectors denoted as $v(x)$ to construct the diffusion tensor.  
We choose values for these parameters according to \cite{mclennan2015}, and such that the mean diffusivity in white matter is $\blue{2\times10^3 \mu \text{m}^2/\text{h}}$ and the mean diffusivity in gray matter is $\blue{1\times 10^3 \mu \text{m}^2/\text{h}}$. The mean diffusivity is the average of the eigenvalues of the diffusion tensor at each voxel in the brain. Therefore, we set $D_w$ to $\blue{6\times 10^3 \mu \text{m}^2/\text{h}}$ and $D_g$ to be $\blue{1\times 10^3 \mu \text{m}^2/\text{h}}$. This effectively treats the assumption of anisotropic diffusion in the brain, where we assume diffusivity happens more efficiently in white matter. 

We also note that $\bm x^k = (x^k(t), y^k(t), z^k(t))$ corresponds to the locations of the $k$-th NSC,  and there are $N(t)$ of those at time $t$ in the simulation. The second term in the reaction-diffusion equation represents chemoattractant internalization by NSC, where the internalization rate is given by $\lambda$. The $N$ cells are assumed to uptake chemoattractants VEGF and MCP-1 according to a normalized Gaussian spatial profile with center at the cell locations, $\bm{x}^k$, and the width of the profile is the cell radius $R$. The internalization is taken to be proportional to the chemoattractant concentration, \blue{$C_j$}. 

\blue{The last term in the equation is the chemoattractant production term $p(\bm{x}, C_j)$. We model chemoattractant production as the sum of a tumor-localized source and a spatially distributed background source of logistic form,
\begin{equation}
p(\bm {x}, C_j) = \chi_T \frac{1}{2}  \left[ 1- \tanh\left(\frac{\| \bm {x} - \bm x_j \|_2 - r}{\omega}\right)\right] (1-C_j) +\chi_{B} C_j(1-C_j)
\end{equation}
where $\bm x_j = (x_j,y_j,z_j)$ is the tumor center, $r$ is the tumor radius, $\omega$ is the width of the transition between the tumor and the surrounding tissue, and $\chi_T$ and $\chi_B$ denote the tumor-associated and background production rates, respectively. The tumor-sourced term is motivated by observations that chemoattractants like VEGF are produced and released at elevated levels within brain tumors \cite{Stockhammer2000VEGF}. The background term represents aggregate VEGF production by nonmalignant cells outside the tumor. VEGF expression has been observed in several regions and cell populations of the mouse brain \cite{Maharaj2006VEGF}. Either production mechanism may be omitted by setting its corresponding coefficient to zero. The coefficient values used for each simulation regime are specified in Section 3.}


Table \ref{tab:param} summarizes the descriptions and values for the parameters used in the hybrid ABM and PDE model. 
\blue{Because direct measurements of NSC migration speed were not available for the present experimental system, the migration parameters were selected by comparing simulated trajectories with the observed long-term migration timescale and refining the values through iterative simulations. We chose $d_w=2 \mu$m/h as a representative effective migration speed in the normal mouse brain. This choice is consistent with experimental observations showing that NSCs remained largely localized near the injection site at 3 months but reached the olfactory bulb from the corpus callosum by 9 months \cite{rockne_long-term_2018}. Moreover, the value lies within the previously reported range of approximately 47–70$\mu$m/day for NSC migration under normal conditions \cite{flexman2011quantitative,chen2015}. Nevertheless, because this parameter was not directly calibrated from time-resolved measurements of individual NSC trajectories, it should be interpreted as an effective model-scale migration speed, and additional longitudinal data will be needed for more rigorous parameter estimation and validation. 
The white-matter threshold $\varepsilon_M$ was selected so that approximately 10\% of the brain domain was classified as white matter, consistent with reported anatomical estimates \cite{zhang2000universal}. Because this threshold determines the spatial extent of tissue-guided migration, variations in $\varepsilon_M$ may substantially affect predicted NSC trajectories and target-arrival efficiencies.}

\begin{table}[h!]
\centering
\caption{Model parameters, their biological interpretation, and their range of values.  }
\begin{tabular}{@{} c l l c @{}} 
\toprule
\textbf{Parameter} & \textbf{Description} & \textbf{Value} & \textbf{Reference} \\
\midrule 
$\epsilon_M$ & Threshold for white matter/gray matter delineation  & 0.4\\ 

$\lambda_{c_1},\, \lambda_{c_2}$ & Chemotaxis sensitivity level & $[0.1, 10]$ \\

$FA(\bm x)$ & Fractional anisotropy representing  degree of anisotropy  & $[0, 1]$ \\
$\bm q(\bm x)$ & The dominant fiber direction resolved from DTI & $\mathbb{R}^3$ \\
$d_g$ & Maximum migration speed of NSC on gray matter & $1 \mu \text{m}/ \sqrt{\text{h}}$  \\ 
$d_w$ & Maximum migration speed of NSC on white matter & $2 \mu \text{m}/\text{h}$ \\
$\beta_{w}$ & Stochasticity in tissue-guided migration & 1  \\
$\alpha_{c}$ & Stochasticity in chemotaxis & 1 \\
$\gamma$ & Chemoattractant internalization rate & $[\blue{100},1000]$/h \\
$R$  & Cell radius (nuclear)  & $7.5 \mu \text{m}$ \\
\blue{$\chi_T$}      & \blue{Tumor-sourced production rate of chemoattractant} & $[0.0001,1]$/h \\
\blue{$\chi_B$}      & \blue{Background production rate of chemoattractant} & $[0.0001,1]$/h \\
$D_w$, $D_g$ & Diffusion in white and gray matter & \blue{$6\times10^3$, $1\times10^3$} $\mu \text{m}^2/\text{h}$\\ 
\hline 
\end{tabular}
\label{tab:param} 
\end{table}

\section{Experimental data and model calibration} \label{sec:data}

\subsection{Experiments of orthotopic brain tumor model and intranasal NSC administration}

\begin{figure}[!htb]
\centerline{
    \includegraphics[width=14cm]{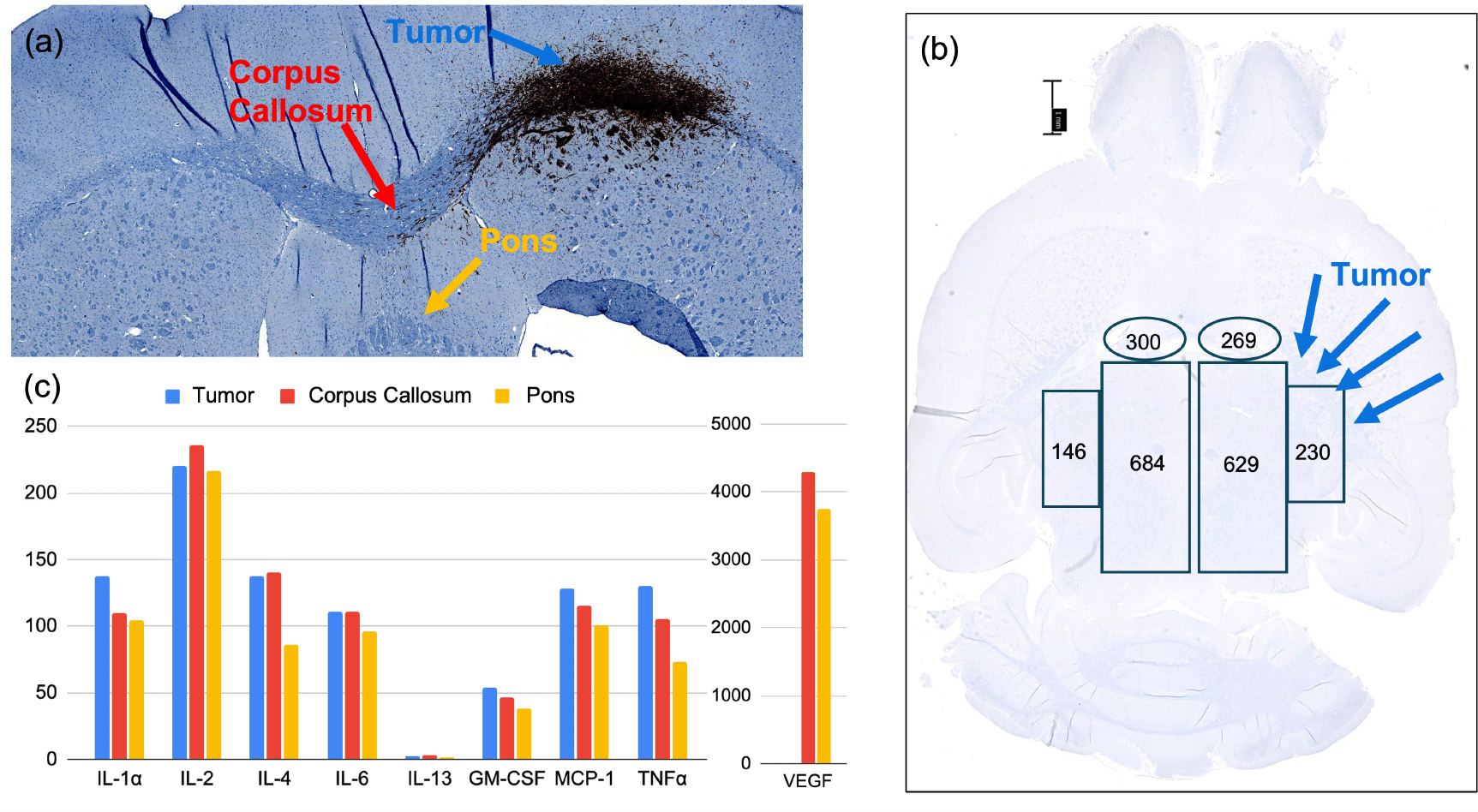}
    }
\caption{ Representative coronal (a) and transverse tissue section (b) indicating tumor location (blue arrows) as well as the box sizes and locations used in subsequent NSC-count analysis. (c) Concentrations of various cytokines measured in vivo in brain slices. 
} 
\label{fig:ExpData}
\end{figure}

Tumor-bearing mice were generated using an orthotopic glioblastoma model. PBT138 brain cancer cells were implanted into the right frontal lobe
on Day 0, producing a localized intracranial tumor. Tumors were allowed to grow for two weeks to permit the establishment of a well-defined tumor mass and its associated microenvironment, where PBT138 human glioblastoma cells formed reproducible intracranial tumors suitable for downstream quantitative analyses of the NSC distribution as shown in Fig. \ref{fig:ExpData}. Beginning on Day 13, therapeutic NSCs, namely LMNSC01 \cite{ojeda2025modeling}, were delivered intranasally, using a repeated-dose regimen consisting of six administrations performed every other day (Days 13, 15, 17, 19, 21, 23). 
To quantify the NSC migration to the tumor site, mice were sacrificed on Day 30, and brains were collected for histological analysis. Using serial sectioning and quantitative cell mapping approaches, the spatial distribution of intranasally delivered NSCs was quantified to assess their degree of infiltration and localization relative to the tumor mass.  
While accumulation of NSCs at the tumor site has been demonstrated across multiple studies in \cite{barish2017,Gutova2019,Gutova2021}, this intranasal experiment revealed that NSC migration did not consistently concentrate around the tumor site. For instance, in one of the mice, NSCs remained broadly dispersed throughout the brain rather than accumulating adjacent to the tumor, as shown in Fig. \ref{fig:ExpData}. The quantified NSC counts showed an approximately even distribution between the left and right hemispheres, despite the tumor being located on the right side. This unexpected pattern raises the possibility that chemotactic cues were insufficient to guide NSCs toward the tumor. Thus, to further investigate why NSCs did not accumulate at the tumor site, we examine cytokines known to mediate chemotactic signaling, with the goal of determining whether insufficient or disrupted \blue{chemoattractant} gradients contributed to the reduced tumor-directed migration. 
To characterize the chemotactic microenvironment encountered by migrating NSCs, we quantified levels of VEGF and MCP-1, two prominent glioma-associated chemoattractants. These cytokines are known to influence NSC tumor-directed migration, consistent with prior observations that NSC migration is regulated by tumor-derived chemokines, growth factors, and other molecular signals \cite{Gutova2008,breton2011impact,kizil2015effects}.

Cytokine data were available at the study endpoint in selected regions of the brain for a subset of mice. Consequently, we approximated the measured \blue{chemoattractant} values $\{c_i\}$ collected at spatial locations $(x_i, y_i, z_i)$ to construct a full \blue{chemoattractant} field over the entire three-dimensional mouse brain. 
Because the sparsity of measurements does not allow reliable reconstruction of a complex spatial profile, we consider two simplified representations of the initial \blue{chemoattractant} profile: Gaussian or constant. 
In the Gaussian case, we assume a unimodal \blue{chemoattractant} profile centered near the tumor, modeling a  chemotactic source localized to the tumor. We approximate the spatial profile with parameters $c_1$, which represents the maximal \blue{chemoattractant} level, and $\sigma$, which represents the variance as 
\begin{equation}
    \bar{c}(x,y,z; c_1, \sigma ) = c_1 \exp \left( - \frac{ (x-x_c)^2 + (y-y_c)^2+ (z-z_c)^2 }{2 \sigma^2} \right), 
    \label{eq:gaussian}
\end{equation} 
where $(x_c, y_c, z_c)$ is the center of the tumor. 
In the constant case, we assume a spatially uniform \blue{chemoattractant} distribution, modeling a scenario in which \blue{chemoattractant} concentrations lack a pronounced gradient. In particular, we define a parameter $c_0$ for a constant concentration as
\begin{equation}
    \bar{c}(x,y,z; c_0) = c_0. 
    \label{eq:const}
\end{equation}
\blue{These profiles were chosen as representative limiting cases, because the available measurements were too spatially sparse to support reliable reconstruction of a more complex three-dimensional chemoattractant landscape.} 
In addition, we include a mean-zero Gaussian field term to account for variability observed in the data. In particular, the \blue{chemoattractant} concentration in the model is initialized as $\bar{c}(x,y,z;\theta^*) + \xi(x,y,z; \omega)$ where $\bar{c}(x,y,z;\theta^*)$ denotes the calibrated deterministic \blue{chemoattractant} profile with parameters $\theta^*$ and $\xi(x,y,z; \omega)$. 

\blue{The two initial profiles are paired with production mechanisms in the reaction-diffusion equation (\ref{PDE}) chosen to maintain their intended spatial organization during the simulation. The Gaussian initial condition is coupled with both tumor-sourced production and weak spatially distributed background production, which sustains an elevated concentration near the tumor while retaining a nonzero background concentration. The approximately uniform regime omits the tumor-sourced production and instead uses logistic production to maintain a more uniform concentration field. These cases should therefore be interpreted as paired initial profile-production regimes representing a tumor-directed gradient and a weak gradient. For the simulations, the Gaussian chemoattractant regime used a tumor-sourced production rate of $\chi_T = 1$/h and a weak background coefficient of $\chi_B = \frac{0.5}{24}$/h. In the constant chemoattractant regime, tumor-sourced production rate is set to zero and $\chi_B = 1$/h. These values were selected to sustain a localized tumor-directed gradient in the Gaussian regime and a comparatively uniform concentration field in the constant regime.}

\begin{figure}[h!]
    \centerline{ \footnotesize (a) \hspace{1cm} }
    \centerline{
    \includegraphics[width=0.7\textwidth]{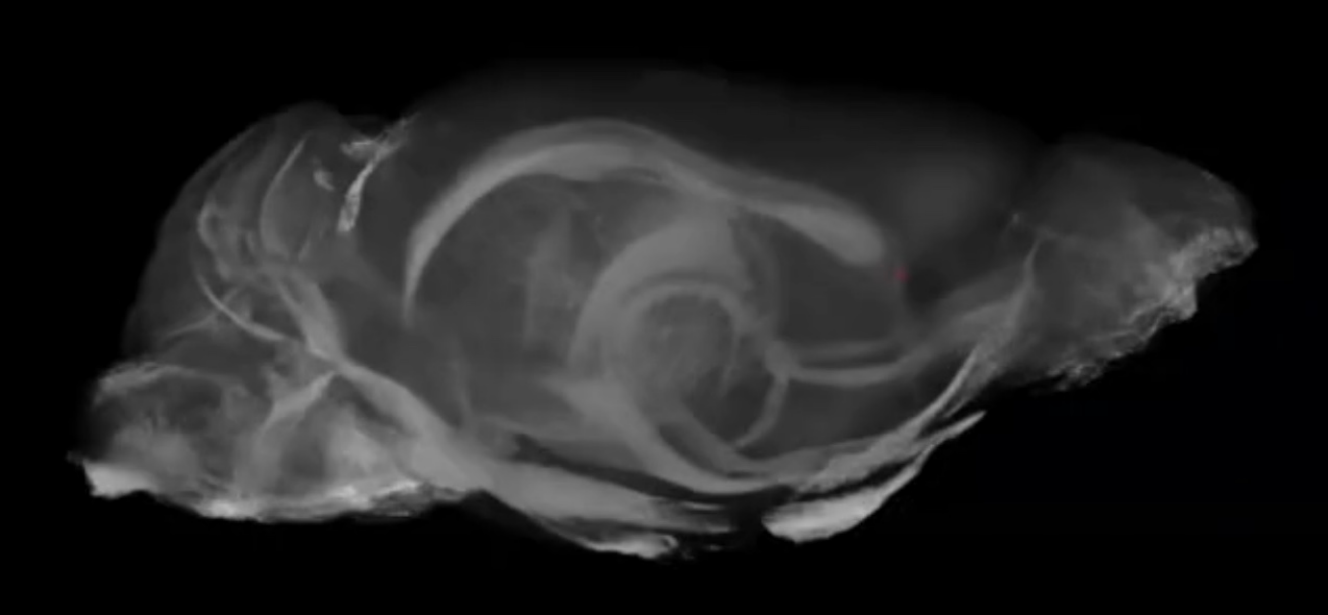} \hspace{1cm}
}
    \centerline{ \footnotesize (b) \hspace{6.5cm} (c) \hspace{1.5cm} }
    \centerline{
    \includegraphics[width=0.48\textwidth]{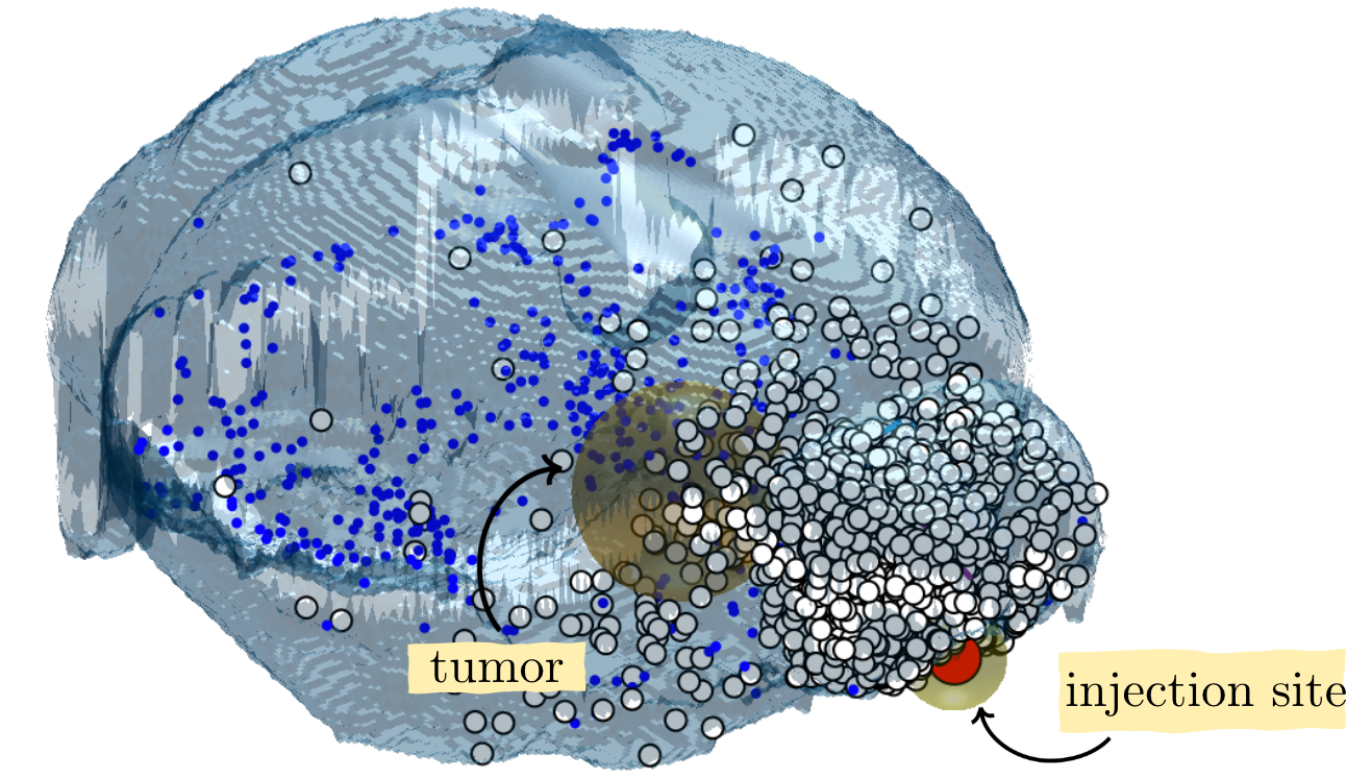}
    \includegraphics[width=0.48\textwidth]{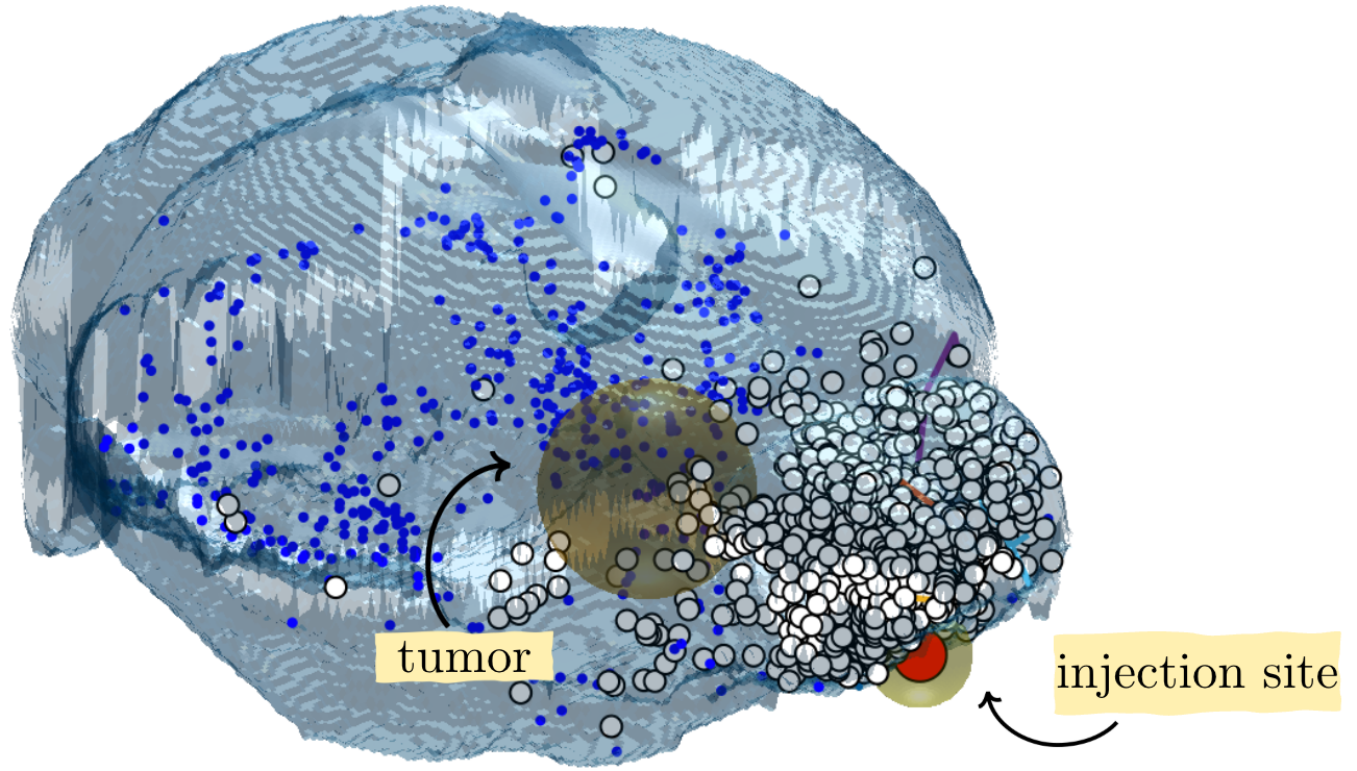}
    }
    \caption{3D mouse brain domain used in our simulation of NSC migration model. \blue{Diffusion tensor imaging (DTI)-derived fractional anisotropy (FA) map of the mouse brain (a), with brighter and darker regions corresponding to local FA values closer to 1 and 0, respectively. The bright, stripe-like regions indicate highly anisotropic white-matter tracts.} Intranasal injection site is colored in red, the tumor in yellow, coarsely sampled white matter tract in blue, and NSCs in white circles. Left image (b) shows cells migrating to the tumor. Right image (c) shows NSC clustered at the injection site.}
    \label{fig:domain}
\end{figure}

\subsection{\blue{Numerical implementation}} 

\blue{The chemoattractant reaction–diffusion Eq. \eqref{PDE} is solved on the irregular three-dimensional brain geometry using the diffuse domain method \cite{Li2009DiffuseDomain}. The physical brain domain is embedded within a rectangular computational domain and represented by a smooth phase-field function, which is approximately one inside the brain and zero outside, with a smooth transition across the brain boundary. This formulation allows the no-flux boundary condition on the brain surface to be incorporated while using a uniform Cartesian grid.

The resulting diffuse-domain equation is discretized using an implicit–explicit time-stepping scheme. The diffusion, internalization, and production terms that contribute to an M-matrix structure are treated implicitly to preserve nonnegativity of the chemoattractant concentration, whereas the remaining cross-diffusion and drift terms are incorporated through a positivity-preserving correction \cite{Kuzmin2005,PLEMMONS1977175}. The computational grid uses \(\Delta x=\Delta y=\Delta z=1\) voxel, and the time step is \(\Delta t=1/200\) days, matching the agent-based model. The resulting sparse linear systems are solved using biconjugate gradient stabilized method with an incomplete-LU preconditioner. Full details of the diffuse-domain reformulation, spatial discretization, and positivity-preserving numerical scheme are provided in Appendix \ref{sec:AppendixA}. 
}

\subsection{Simulation of NSC migration under constant and Gaussian \blue{chemoattractant} profiles}

Here, we present the results of model calibration on the 3D mouse brain domain as in Fig. \ref{fig:domain} under the two assumed \blue{chemoattractant} profiles--Gaussian in Eq. \eqref{eq:gaussian}  and constant in Eq. \eqref{eq:const}--each of which yields a distinct parameter regime consistent with the experimental observations. These two scenarios define feasible regions of parameter values, particularly for the chemosensitivity parameter $\lambda_c \in [1, 10]$, that can reproduce the observed NSC distributions.

In the Gaussian \blue{chemoattractant} case, the chemotactic field exhibits a single peak centered near the tumor. In this setting, successful tumor-directed migration occurs only when the chemosensitivity parameter is large enough.  
When the chemosensitivity is $\lambda_c \geq 4$, NSCs successfully arrive at the tumor site as shown in Fig. \ref{fig:domain}(a). For the NSCs to fail to accumulate at the tumor site, the chemosensitivity has to be small. In this scenario, the \blue{chemoattractant} gradient is insufficient to overcome tissue-guided motility, leading again to poor tumor arrival. Therefore, calibration to the experimental data identifies a narrow feasible interval of sufficiently small chemosensitivity values $\lambda_c \leq 1$ that matches the observed experimental distribution.

In the constant \blue{chemoattractant} case, the concentration of chemotactic cues is spatially uniform across the brain. Under this assumption, the model predicts that variations in the chemosensitivity parameter have little to no effect on NSC localization. This is because a uniform \blue{chemoattractant} field generates no directional gradient. Consequently, NSCs disperse broadly rather than accumulating at the tumor site as shown in Fig. \ref{fig:domain}(b), consistent with the experimentally observed hemispherically symmetric distribution.

Taken together, these results suggest two possible explanations for the insufficient NSC accumulation at the tumor site. Under the uniform \blue{chemoattractant} scenario, the failure of NSCs to localize to the tumor arises because the \blue{chemoattractant} field lacks a chemotactic gradient. Under the Gaussian scenario, the NSC localization failure arises from small chemosensitivity, either too weak to detect the \blue{chemoattractant} peak or too noisy, causing NSCs to localize at distinct locations before reaching the tumor. These mechanisms motivate combination strategies aimed at modulating \blue{chemoattractant} levels or NSC responsiveness to enhance therapeutic targeting by altering the \blue{chemoattractant} profile in the tumor and surrounding brain.

\section{\blue{Simulation of combination treatment-induced perturbations}}\label{sec:result}

We investigate three combination strategies designed to enhance NSC migration to the tumor site by modulating NSC responsiveness or \blue{chemoattractant} distributions. The strategies include IL-4 or IL-10 pretreatment, CXCL12 pretreatment, and radiation therapy. Each strategy targets a different component of cell migration and chemotactic microenvironment and aims to overcome the limitations identified in the previous section. 
\blue{Rather than explicitly modeling the full biological mechanisms of each combination treatment, we investigate experimentally motivated downstream effects through perturbations of model parameters and chemoattractant dynamics. These simulations therefore serve as mechanistically motivated sensitivity studies to determine how treatment-induced changes in NSC responsiveness, motility, and chemotactic signaling may influence tumor-directed migration.}

\begin{table}[h!]
\centering
\caption{Three combination treatments to improve NSC migration to the tumor site.}
\begin{tabular}{@{} c l l c @{}} 
\toprule
\textbf{Combination treatment} & \textbf{Description} & \textbf{Model} & \textbf{Reference} \\
\midrule 
IL-10 &  increased fraction of migrating NSCs  & increase $\alpha_c$ & \cite{guan2008} \\ 
CXCL12 &  increased migration speed  & increase $d_w$ & \cite{chen2015} \\ 
Radiation therapy & \blue{chemoattractant} profile fluctuation   & increase $\nabla C(t,x)$ & \cite{tabatabaei2017radiotherapy} \\ 
\hline 
\end{tabular}
\label{tab:combination} 
\end{table}

\subsection{Pretreatment with anti-inflammatory cytokine (IL-10) }
Anti-inflammatory cytokines can facilitate NSC migration by stabilizing and protecting the neural micro-environment \cite{livingston2023neural,boltze2020increased}. In particular, Guan et al. \cite{guan2008} studied the effect of anti-inflammatory cytokines, IL-10 and IL-4, on chemotactic response of NSC to the chemokines. The authors showed that exposure of NSCs with IL-4 or IL-10 significantly increases the expression of several chemokine receptors, including CCR1, CCR2, CXCR4, and CXCR7, known to regulate NSC migration. This resulted in increased chemotactic response to chemokines such as SDF-1 and Rantes. Compared to the control condition, in which only 4–5\% of the NSCs reached the tumor, pretreatment with IL-10 increased migration to 7–15\% in a dose-dependent manner.

We therefore incorporate IL-10 pretreatment in our model by increasing the stochastic chemosensitivity parameters given as $\lambda_c$ times $\eta(w)$. $\lambda_c$ is the maximum chemosensitivity value which takes values between 1 through 10. However, in the Gaussian case, the feasible parameter range is restricted to low values near one. The stochasticity parameter $\eta(w) \sim \blue{\textrm{Beta}(\alpha_c, 1)}$ determines the proportion of NSCs that migrate further. Our default value $\alpha_c=1$ makes $\eta(w)$ uniformly distributed on $[0,1]$, and larger values of $\alpha_c$ make the distribution skewed toward $1$, which will have more NSCs sensitive to the \blue{chemoattractant} gradient. We test values for $\alpha_c$ larger than $1$ to model the effects of IL-10 pretreatment of the NSCs. Specifically, we take $\alpha_c=[1,2,4,8,16]$ for each chemosensitivity value. As $\alpha_c$ increases from one, more NSCs have larger chemosensitivity magnitudes, and the chemosensitivity term becomes more deterministic. 

\begin{figure}[!htb]
\centerline{
    \includegraphics[width=14cm]{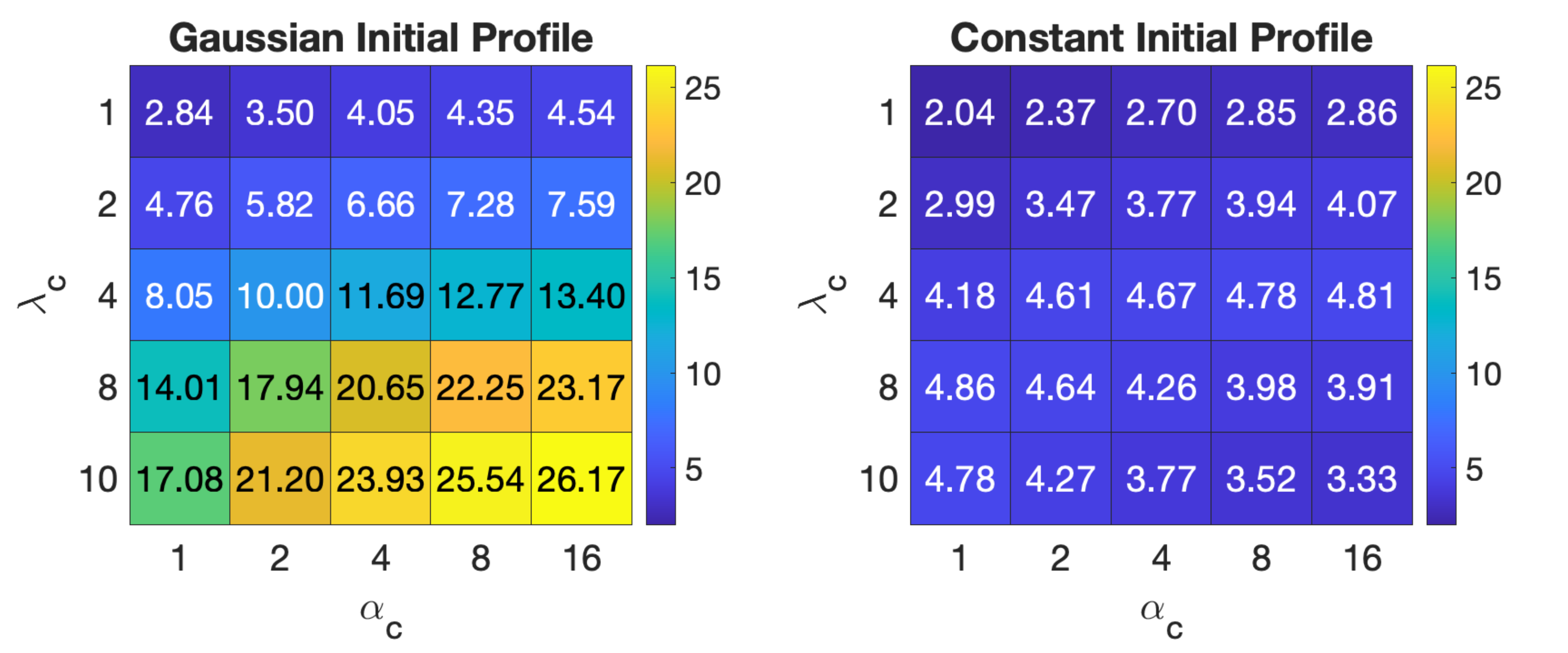}}
\caption{  
\blue{Average} NSC arrival percentage at the tumor for a range of chemosensitivity $\lambda_c$ and $\alpha_c$ values for Gaussian (left) and constant (right) profile for \blue{chemoattractants} $C(t,x)$. Values are averages across ten simulations. $\lambda_c$ is the maximum chemotactic sensitivity level and $\alpha_c$ parameterizes the stochastic chemotactic sensitivity parameter via $\blue{\text{Beta}}[\alpha_c, 1]$.
}
\label{fig:heatmapalpha}
\end{figure}

In the Gaussian profile case, there is a clear response to an increase in \blue{$\lambda_c$ and} $\alpha_c$. The \blue{average} percent on cancer increases, which is \blue{more} notable when chemosensitivity values are on the \blue{moderate-to-}larger end of the \blue{range considered} as shown in Fig. \ref{fig:heatmapalpha}. We see over a \blue{9}\% increase of tumor arrival as $\alpha_c$ increases from 1 to 16 when chemosensitivity is \blue{equal to or above} 8. However, within the low-chemosensitivity regime, identified as feasible from our experimental data, increasing $\alpha_c$ produces only a moderate effect, resulting in a marginal improvement in tumor arrival. 

In the constant profile case, it was observed that the \blue{average} percent on cancer over time \blue{does not change in a notable way with increased} $\lambda_c$ and $\alpha_c$. This result is expected because NSCs do not have a \blue{significant} chemotactic gradient to follow \blue{as in the Gaussian profile case}, as the concentration is roughly constant \blue{over time}, \blue{so increasing these parameters does not benefit arrival rates}. Consequently, their movement is governed primarily by tissue-guided motility and stochastic fluctuations rather than chemotaxis.
The results of this can be seen in Fig. \ref{fig:heatmapalpha}, where the heat map for the constant profile shows \blue{minor fluctuations in} tumor arrival percent across \blue{increasing} parameter regimes compared to the scale of values seen in the Gaussian profile case.

From this, we infer that IL-10 pretreatment is ineffective under a constant \blue{chemoattractant} profile and moderately beneficial under a Gaussian profile when chemosensitivity is low ($\lambda_c = 1$), indicating that it is unlikely to improve NSC tumor arrival in our experimental mouse model. However, for patients whose tumors generate sufficiently strong chemotactic cues, IL-10–mediated enhancement of chemokine receptor expression could substantially improve NSC migration to the tumor site.

\subsection{Pretreatment with chemokine (CXCL12) }

Pretreatment of neural progenitor or stem cells with chemotactic factors is known to enhance their migratory abilities. In Chen at el. \cite{chen2015}, CXCL12 (SDF-1) and CD184 have been shown to upregulate its cognate receptor CXCR4 in neural progenitor cells and strengthen chemotactic response, and they show that continuous administration of CXCL12 along the corpus callosum resulted in a dramatic improvement in neural cell migration as transplanted cells migrated up to 1,881$\mu$m in 7 days compared to only around 200$\mu$m without CXCL12. The average migration speed increased by approximately 9 times the initial speed under CXCL12 exposure compared to the control case,  and the volume of the migratory path expanded nearly 30-fold. These findings demonstrate that improving CXCL12/CXCR4 signaling can substantially accelerate and extend neural cell movement along white-matter tracts.

\begin{figure}[!htb]
\centerline{\footnotesize \textsf{(a)} \hspace{4.3cm} \textsf{(b)} \hspace{4.7cm} \textsf{(c)} }
\centerline{
    \includegraphics[width=0.3\textwidth]{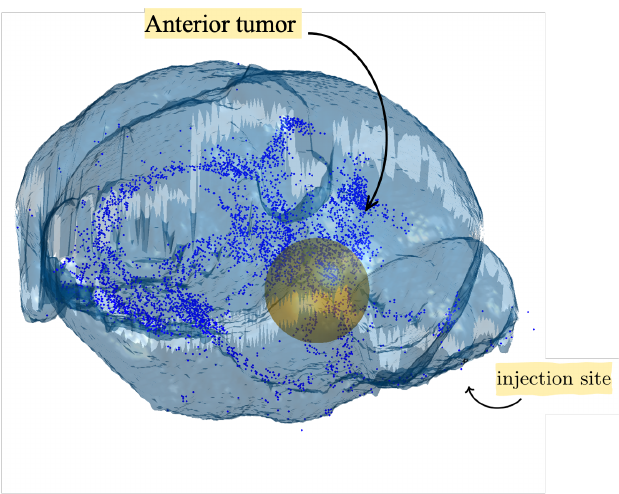}
    \includegraphics[width=0.7\linewidth]{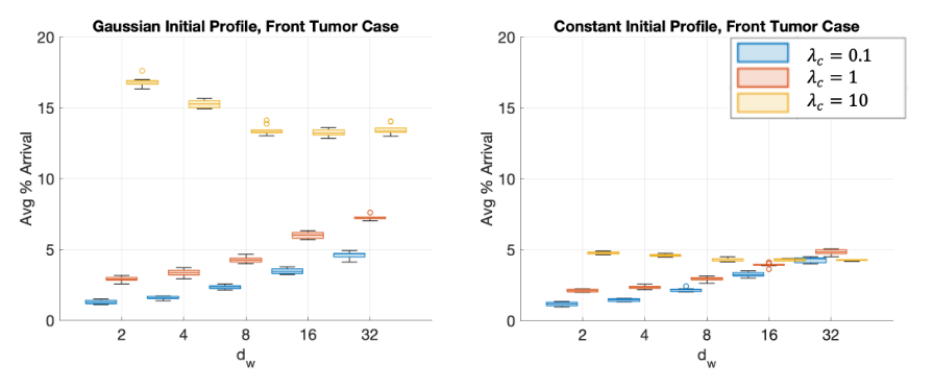}
}
\centerline{\footnotesize \textsf{(d)}  \hspace{4.3cm} \textsf{(e)}  \hspace{4.7cm} \textsf{(f)} }
\centerline{
    \includegraphics[width=0.3\textwidth]{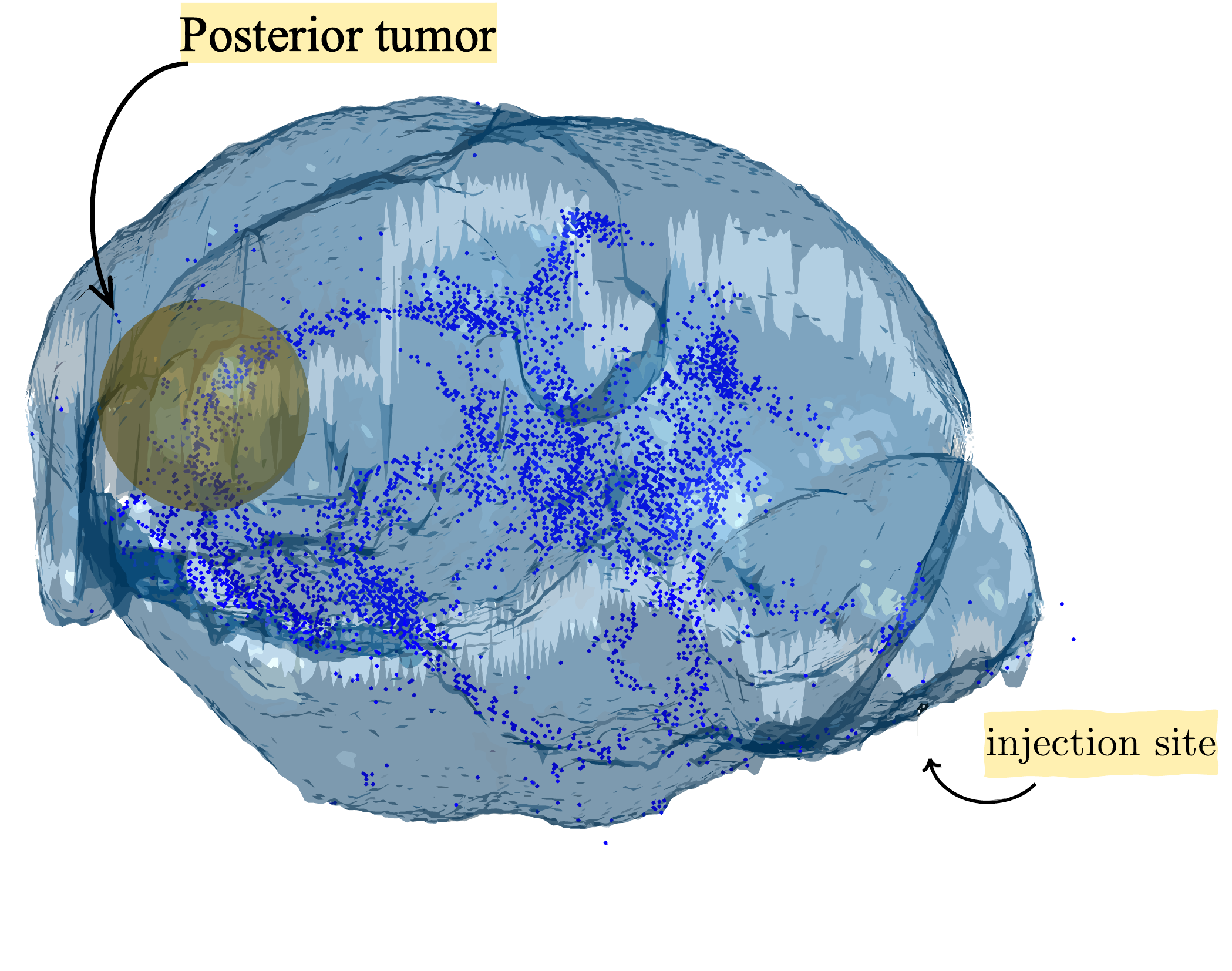}
    \includegraphics[width=0.7\linewidth]{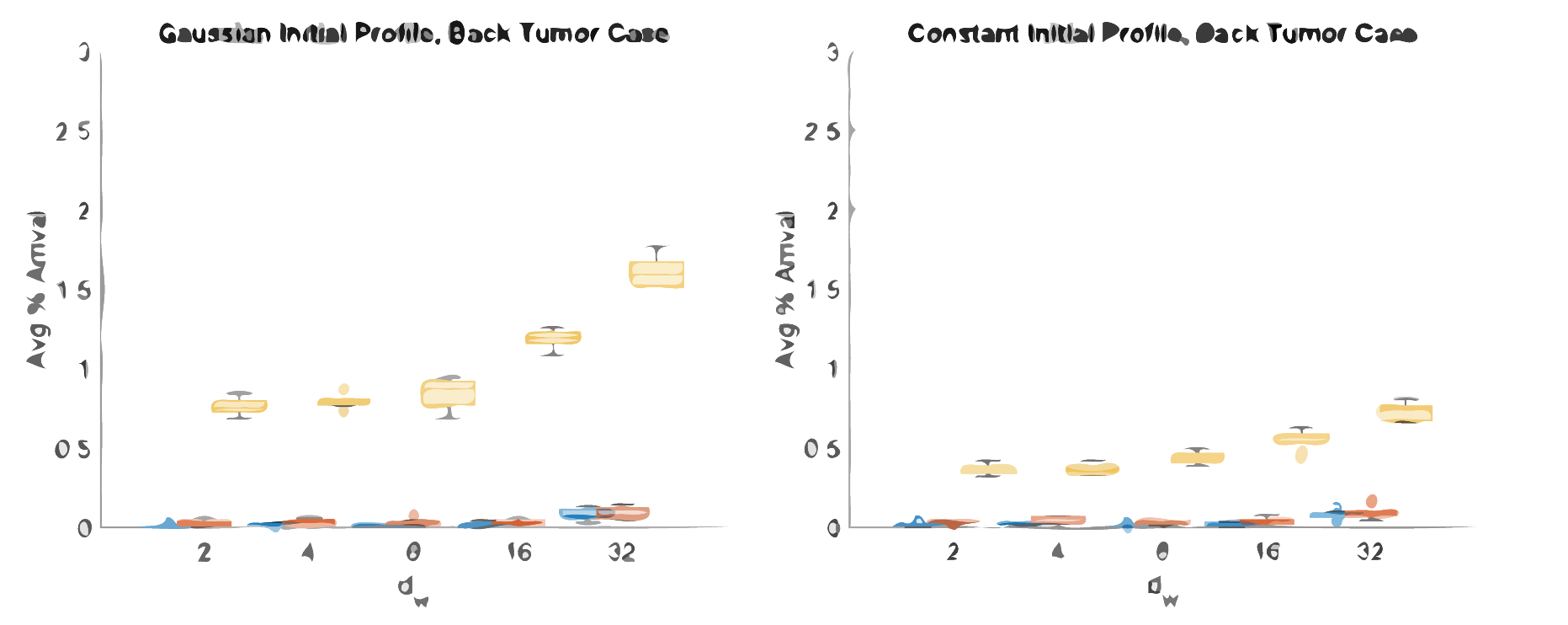}
}
\caption{ The \blue{average} percent arrival of the NSC at the tumor for increased white matter migration step $d_w = 2, 4, 8, 16, 32 $ and chemosensitivity $\lambda_c =$ 0.1, 1, or 10. Shown are the tumor locations (yellow) and results at the anterior brain (a,b,c) and the posterior brain (d,e,f) under the Gaussian \blue{chemoattractants} profile (b,e) and constant \blue{chemoattractants} profile (c,f). 
}
    \label{fig:dotplotfront}
\end{figure}

Motivated by these results, we incorporate CXCL12-based modulation into our simulation by increasing $d_w$, the migration step size of the cells on white matter. We simulate $d_w$ values 2, 4, 8, 16, and 32, where the 9-fold increase shown in \cite{chen2015} falls between 16 and 32.   
We test for chemosensitivity values $\lambda_c = 0.1, 1$ and $10$ for VEGF, and two scenarios considering the tumor location, when the tumor is either in the anterior brain or the posterior brain, as shown in Fig. \ref{fig:dotplotfront}(a, c).

Fig. \ref{fig:dotplotfront} shows the \blue{average} tumor arrival across 10 simulations. When the tumor is in the anterior region of the brain, closer to the injection site, increasing the step size in white matter improved the percent of tumor arrival under the Gaussian \blue{chemoattractant} profile \blue{for lower chemosensitivity values}. We observe this effect because the increased tissue-directed migration speed allows NSCs to explore a larger region of the brain and enables them to reach locations where they can detect the chemotactic gradient, which directs them toward the tumor. In particular, for low chemosensitivity values, $\lambda_c = 0.1$ and $1$ that agree with the unsuccessful experimental outcomes, increasing the migration speed in white matter doubled the percent of NSCs arriving at the tumor site. \blue{However,} when the chemosensitivity is   $\lambda_c = 10$, the impact of larger $d_w$ was \blue{diminishing}. \blue{This could be explained by a significant number of cells already reaching the tumor for small $d_w$, so the increased motility results in cells migrating beyond the tumor site.} On the other hand, with the constant \blue{chemoattractant} profile, larger values of $d_w$  modestly increased the percent of NSCs at the tumor site \blue{when chemosensitivity is lower} due to NSC's capability to explore larger regions of the brain, \blue{similar to the Gaussian profile case}. \blue{When chemosensitivity was large, the diminishing effect was minimal, which could be explained by the cells not being greatly influenced towards the tumor by strong gradients. Therefore, a smaller number of cells reach then migrate beyond the tumor site with increased $d_w$ compared to what was seen in the Gaussian profile case. } 


We further explored the scenario when the tumor is located in the posterior region of the brain. Our previous work in \cite{gomez_mathematical_2022} showed that  intranasal injection may be ineffective when the tumor is too far away from the injection site, and it reduces the chance that NSCs will arrive at the  tumor. Here, we study whether CXCL12 pretreatment can help with this case. Fig. \ref{fig:dotplotfront}(c,d) suggests that it can help under the Gaussian \blue{and constant }profiles and with strong chemosensitivity. When chemosensitivity is large, meaning in our simulation $\lambda_c = 10$, under the Gaussian \blue{and constant profiles}, increasing $d_w$ \blue{nearly} doubles the tumor arrival percent. \blue{This can be explained by the heightened sensitivity of the cells to gradients and their increased mobility. In these cases, we observe similar trends but different magnitudes of arrival numbers, which could be a result of the cells creating their own gradient to make it near the tumor site and benefitting from the increased step size on white matter. Once the stronger gradient is detected at the tumor in the Gaussian profile case, more cells can be directed to the tumor compared to the constant profile case, where the stronger gradient would not be as distinctly formed.} \blue{In the lower chemosensitivity regimes}, the outcome remains unchanged.

Overall, our model demonstrates that CXCL12 pretreatment can be effective in improving the tumor arrival, especially under a Gaussian profile for the \blue{chemoattractant} distribution by enhancing the NSCs ability to explore broader regions of the brain. In particular, the pretreatment doubles the tumor arrival when chemosensitivity is low, the conditions matching our unsuccessful experiments. 

\subsection{Tumor radiation}

In our simulations, IL-10 and CXCL12 pretreatment showed a small benefit under the constant \blue{chemoattractants} profile, demonstrating a fundamental limitation when the chemotactic landscape does not have a meaningful gradient, and enhanced chemosensitivity or motility cannot drive NSCs toward the tumor. To improve directional cues towards the tumor, an intervention to actively reshape the \blue{chemoattractants} environment is needed, which involves creating or strengthening gradients that can be detected and followed. This motivated our exploration of radiotherapy, which is known to alter the \blue{chemoattractants} landscape by inducing inflammatory and stress responses \cite{liu2023radiotherapy,merchant2009cytokine}. Levels of pro-inflammatory cytokines and chemokines are elevated due to DNA damage, and therapeutic cells and immune cells are recruited to contribute to tissue repair and immune modulation. Thus, radiotherapy has the potential to generate stronger chemotactic gradients that can direct NSC migration more effectively. 
In particular, radiation has been shown to increase the levels of inflammatory cytokines, such as MCP-1, with each added dose \cite{tabatabaei2017radiotherapy}. 
\blue{Thus, we simulate the radiotherapy-induced changes in the chemoattractant environment.}

 We consider VEGF and MCP-1 in this simulation and assume, as before, that our \blue{chemoattractant} profile is either a constant or a Gaussian profile. Moreover, we assume that the chemosensitivity parameters for VEGF and MCP-1 are the same value. Different dosing schedules are compared, considering the administration of 1, 3, or 5 radiation fractions/doses. With each added dose of radiation, we increase the levels of MCP-1 locally at the tumor site according to \cite{tabatabaei2017radiotherapy}. In particular, the ratio of increase from the baseline MCP-1 level to the first, third, and fifth dose levels are  $1.7$, $1.0$, and $1.3$, respectively. 

In addition to modifying the concentration gradients in response to irradiation, we also acknowledge and account for the treatment resulting in the death of NSCs locally at the tumor site. We assume NSC death can occur at the tumor according to the linear-quadratic model of radiotherapy \cite{Enderling,mcmahon2018linear}, given as follows:
\begin{equation}
    S = e^{-\alpha d -\beta d^2 }, 
    \label{eq:LQ} 
\end{equation}
where $S$ is the probability of cell survival after one fraction of radiation with dose $d$ in units Gray (Gy). We choose the same dose level $d=2$ Gy as in \cite{tabatabaei2017radiotherapy}, and radiosensitivity parameters $\alpha=0.1 \ \text{Gy}^{-1}$ and $\beta = 0.03 \ \text{Gy}^{-2}$ \cite{barazzuol2010mathematical}. The choice for $\alpha$ reflects the moderate radiosensitivity of the NSCs, and the choice for $\beta$ ensures cumulative effects of increased dosage on NSC death. We note that modifications to MCP-1 concentration are only made when 1,3, and 5 radiation fractions are delivered, following the increase pattern seen in \cite{tabatabaei2017radiotherapy}. To model the stochastic radiation-induced NSC death, we apply the survival probability stated in Eq. \eqref{eq:LQ} to determine if the cells at the tumor site survive or not when radiation is applied. In particular, during the treatment period, a NSC that arrives at the defined tumor radius is probabilistically eliminated if a generated uniform random number exceeds the computed survival probability, $S$. 
We run the simulation for 17 days, where the radiation treatment is applied on days 4 through 8.

\begin{figure}[!htb]
\centerline{
    \includegraphics[width=14cm]{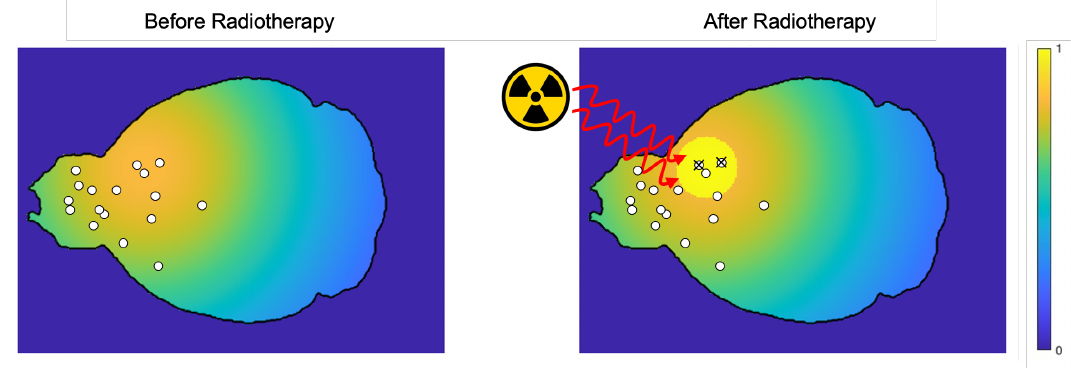}}
\caption{  (Left) Pre-radiotherapy levels of MCP-1 concentration cross-section that is higher at the tumor site and NSC in white circle shown as an example. (Right) Post-radiotherapy levels of MCP-1 concentration shows a localized increase at tumor, with warmer colors indicating higher MCP-1 levels. Crosses indicate NSCs that are removed stochastically at the tumor site due to radiation.
}
\label{fig:MCP-1increase}
\end{figure}

\begin{figure}[!htb]
\centerline{
    \includegraphics[width=1\linewidth]{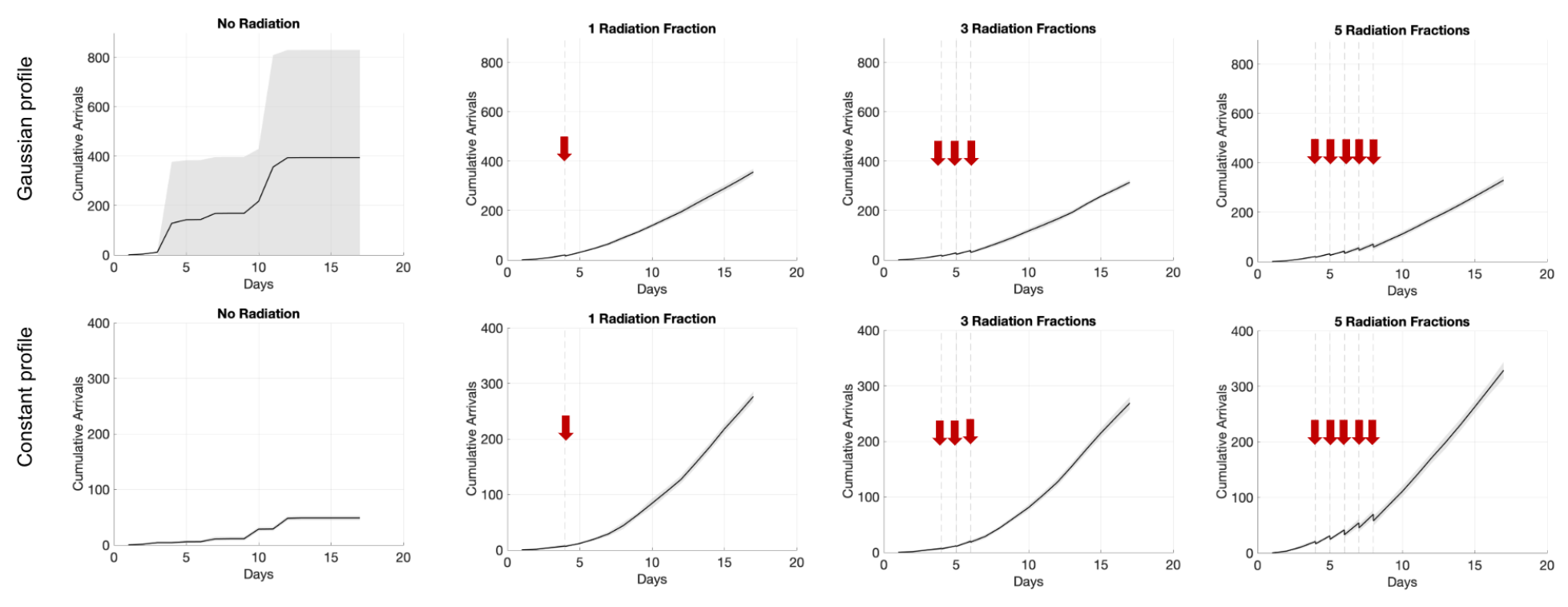}
    
}

\caption{ Number of NSCs arrived at the tumor site under Gaussian (top) and constant (bottom) \blue{chemoattractants} profile. The black line is the mean and the gray area is one standard deviation. Radiation is delivered in 1, 3, or 5 equal fractions on day 4, days 4–6, or days 4–8, respectively, as indicated by red arrows.
}
    \label{fig:arrivalnumradio}
\end{figure}

Fig. \ref{fig:arrivalnumradio} shows the cumulative arrival at the tumor site when the chemosensitivity parameter is \blue{$\lambda_c = 1$}, which is where we find a measurable difference between the dose simulations. The black line is the mean trajectory, and the shaded area is within one standard deviation. 
For the Gaussian \blue{chemoattractant} profile case, increasing the number of radiation doses \blue{did not substantially improve} NSC arrival at the tumor. \blue{However, the added radiation fractions resulted in more consistent NSC arrival numbers compared to the zero-dose case, which had lower arrival numbers for many simulation days overall compared to the one, three, and five-dose radiation treatment plans. A few trials in the zero-dose case had comparatively large NSC arrivals in contrast to the other trials for the same day. For this reason, the standard deviation is much larger in the zero-dose case compared to the other dose plans.} This \blue{benefit of more consistent larger NSC arrivals} is because additional radiation boosts chemoattractant gradients, while cell death due to radiation remains relatively small compared to increased directed migration. For the constant \blue{chemoattractant} profile case, radiotherapy increased the level of NSC arrival compared to no radiation. However, different dose plans produced similar results, with the \blue{five}-dose schedule yielding the highest-arrival percentage on average. Thus, coupling with radiotherapy improves the outcome under a constant \blue{chemoattractants} profile, but additional doses give little added value. In this case, a smaller number of radiation doses \blue{could be considered most beneficial}, especially considering the potential negative side effects of radiation alongside the radiation-induced NSC death.

\section{Conclusion and Discussion}\label{sec:conclusion}

In this work, we developed a hybrid agent-based and continuum model to study intranasal NSC delivery for brain cancer and to evaluate combination strategies for improving tumor targeting. 
Our work suggests  that the selection of combination therapies to enhance tumor arrival of NSCs can be sensitive to  \blue{chemoattractants} profiles in the tumor and surrounding brain. 
By considering both Gaussian and spatially uniform profiles of VEGF and MCP-1 chemo-attractants, we showed that IL-10 pretreatment enhances NSC homing to the tumor only when localized chemotactic gradients are present. Additionally, CXCL12 pretreatment improves migration by enabling NSCs to explore larger regions of white-matter tracts, particularly for tumors located in the anterior region of the brain, distant from the site of NSC administration. Radiotherapy can be an effective strategy by reshaping the chemoattractant landscape, significantly improving NSC arrival under uniform profiles. Our work suggests that the efficacy of treatment is strongly dependent on the underlying \blue{chemoattractants} profile as well as NSC responsiveness to chemotactic cues, which may be actively modulated to increase the probability of NSC migration to the tumor site. 
Table~\ref{Table3} provides a summary of our results across \blue{chemoattractants} profiles and chemosensitivity regimes.

\blue{
Future explorations of this work could be done by relaxing the idealized assumptions imposed on the \blue{chemoattractants spatial} profile as more detailed experimental data becomes available. In particular, our current work assumes simplified chemoattractant distributions, Gaussian or uniform profiles, due to limited spatial measurements that were too sparse to support reliable reconstruction of a more complex three-dimensional chemoattractant landscape. However, chemoattractant distributions are likely to be spatially heterogeneous, time dependent, and shaped by tumor secretion, tissue anisotropy, cellular uptake, degradation, and treatment-induced changes. As richer spatial and temporal datasets become available—for example, through regionally resolved ELISA measurements, longitudinal assays, spatial transcriptomics, or molecular imaging—the model can be calibrated to more realistic chemoattractant fields and used to quantify uncertainty in their spatial dynamics. Incorporating these data will improve the predictive accuracy of the model and support more individualized optimization of NSC delivery and combination treatment strategies.
}

\blue{Our current work includes VEGF and MCP-1 because spatial measurements of both chemoattractants were available from our experimental collaborators. However, the available data were insufficient to estimate chemoattractant-specific chemosensitivity parameters or determine their relative magnitudes. Moreover, VEGF and MCP-1 may influence migration through distinct and interacting mechanisms. VEGF is an established chemoattractant for neural stem and progenitor cells and acts in long-range chemotaxis, whereas MCP-1 can directly promote short-range chemotaxis and has also been shown to enhance VEGF production in some tumor contexts  \cite{lien2020monocyte}. These suggest that the two signals may act over different spatial scales. Future studies incorporating chemoattractant-specific migration assays and more detailed spatial and temporal concentration measurements will be needed to estimate these parameters independently and characterize interactions between VEGF and MCP-1.}
Also, our current model includes only VEGF and MCP-1 as representative chemoattractants. However, NSC migration is modulated by a broader set of cytokines. Thus, we propose to expand our model by coupling additional key regulators, such as CXCL12 \cite{chen2015} and IL-6\cite{islam2009interleukin,kummer2021role}, that are known to affect NSC migration. Including these cytokines will allow our model to simulate complex cytokine signaling environment reflected in tumor microenvironment. This will naturally lead to incorporating signaling and gene regulatory networks to the NSC agent based model, as the current model only includes chemokine consumption. Incorporating mechanistic cell signaling network to NSC agents to enable phenotypic switching, receptor expression, and signal-dependent transitions, will allow a more accurate prediction of NSC variability and downstream transcriptional responses of the pretreatment strategies. 

\blue{In addition, another draw back comes from the fact that direct measurements of NSC migration speed were not available for the present experimental system. While the values are selected to match the experimentally observed long-term migration timescale, nevertheless, because this parameter was not directly calibrated from time-resolved measurements of individual NSC trajectories, it should be interpreted as an effective model-scale migration speed, and additional longitudinal data will be needed for more rigorous parameter estimation and validation. 
Another important quantity to be studied is the white-matter threshold $\varepsilon_M$. While it was selected to match the percentage of white matter, because this threshold determines the spatial extent of tissue-guided migration, variations in $\varepsilon_M$ may substantially affect predicted NSC trajectories and target-arrival efficiencies. Further sensitivity analysis } 
Finally, our simulations were calibrated to murine glioblastoma data, but the mathematical model is readily extendable to human-scale applications using patient-derived imaging data. 
\blue{However, although this model can be readily adapted to human-specific brain anatomy and tractography, the present quantitative results are specific to the mouse brain. The substantially larger size of the human brain, together with differences in white-matter architecture, chemoattractant transport, migration length scales, and NSC behavior, may alter migration times, target-arrival efficiencies, and treatment responses. Therefore, application to humans will require thorough species-specific scaling, parameter estimation, and validation using human or clinically relevant experimental data.}
For instance, our hybrid model can be embedded with three-dimensional human brain reconstructed from MRI and DTI-based tractography. Our future work includes in silico studies across anatomical variability, tumor locations, and delivery routes, to study personalized NSC therapies in the human brain.

\begin{table}[h!]
\centering
\begin{tabular}{lcccc}
\hline
\textbf{Combination treatment} &
\multicolumn{2}{c}{\textbf{Effect in constant profile}} &
\multicolumn{2}{c}{\textbf{Effect in Gaussian profile}} \\
\cline{2-3} \cline{4-5}
& $\lambda_c$ small & $\lambda_c$ large
& $\lambda_c$ small & $\lambda_c$ large \\
\hline
IL-10              &  Low & Low & Moderate     & High \\
CXCL12 (anterior brain)     & Moderate  &  \blue{Low} & Moderate     & \blue{Low} \\
CXCL12 (posterior brain)      & Low & \blue{Moderate}   & Low      & Moderate \\
Radiotherapy       & High &  High & \blue{Moderate}     & \blue{Moderate} \\
\hline
\end{tabular}
\caption{Effects of combination treatments with constant and Gaussian \blue{chemoattractant} profiles depending on chemosensitivity $\lambda_c$.}
\label{Table3}
\end{table}


\appendix

\section{\blue{Numerical implementation of the chemoattractant reaction-diffusion equation}}\label{sec:AppendixA}

\blue{The reaction diffusion equation that models chemoattractant dynamics in our coupled model is given as \eqref{PDE}. Since this equation is defined on irregular geometry, we use the Diffuse Domain Method \cite{Li2009DiffuseDomain} to solve it. This method embeds the original domain into a rectangular computational domain, where standard discretization techniques can be used. This is done with the help of a phase field function, $\phi$, which is given as
$$ \phi(x) =  \frac{1}{2}\left(1-\tanh{\frac{{3r(x)}}{\epsilon}}\right)$$
where $r(x)$ is the signed distance function to the brain boundary. The phase field function is approximately $1$ inside the brain and approximately $0$ outside, with a smooth transition across the boundary. In the implementation,  we take $\epsilon$ to be two voxels. The diffuse-domain reformulation of \eqref{PDE} with Neumann zero boundary conditions on the physical brain boundary is 
\begin{equation}\label{DDMPDE}
    (\phi \ C_j)_t = \nabla \cdot (\phi \ \mathbf{D} \ \nabla C_j) \ + \nabla \cdot(\phi (\nabla^T\mathbf{D})C_j) - \phi  \ C_j \sum_{k=1}^{N(t)} \frac{\gamma}{(2\pi R^2)^{\frac{3}{2}}}\text{exp}\left(-\frac{\| \bm x - \bm x^k \|_2^2 }{2R^2}\right) + \phi \ p(\bm{x},C_j)
\end{equation}
The physical brain domain is embedded in a larger rectangular computational domain, with zero voxels padded around this domain, to ensure the boundary of the brain is away from the boundary of the computational domain. Periodic boundary conditions are assumed on the outer boundary of the computational domain, and the no-flux condition on the original domain is accounted for through our reformulation with $\phi$. To solve this reformulated problem, we implement an implicit-explicit scheme for stability purposes, using the Backward Euler method. The time step matches that of the agent-based model which is 200 steps for each day. Therefore, we set $\Delta t = 1/200$ days. We discretize the domain using a uniform Cartesian grid with $\Delta x=\Delta y=\Delta z=1$ voxel. In the implementation, we evolve the $\phi$-weighted concentration $ q_j = \phi C_j$ rather than $C_j$ directly. The concentration is recovered after each update by $C_j = q_j/\phi$.

Because $C_j$ is the chemoattractant's concentration, a physical quantity, the numerical method should preserve positivity. The positivity assurance relies on the structure of what is called an M-matrix, which for our problem is the backward Euler matrix. Following \cite{Kuzmin2005} a nonsingular matrix $A = a_{ij}$ is an M-matrix if its off-diagonal entries satisfy $a_{ij} \leq 0, \  \text{where} \ \ i\neq j$ and its inverse is entrywise nonnegative, $A^{-1} \geq 0$. In practice, this property can be verified by showing $A$ has nonpositive off-diagonal entries and is diagonally dominant \cite{PLEMMONS1977175}. Our discretized system can be written as a matrix equation, $Aq_j=b$ with $A$ an M-matrix and $b\geq 0$, then $q_j = A^{-1}b \geq 0$. This property motivates how we handle the terms in Equation \eqref{DDMPDE}. 

The diagonal diffusion terms, involving $D_{11}, D_{22}, D_{33}$ are discretized using a finite-volume, face-averaged scheme, which results in a conservative stencil with nonnegative off-diagonal entries and negative diagonal entries. The resulting matrix will become part of the backward Euler matrix since it satisfies the conditions to maintain positivity of the solution \cite{Kuzmin2005,PLEMMONS1977175}.

The off-diagonal, or cross-diffusion, terms involving $D_{12}, D_{13}, D_{23}$ are treated separately. Since $\mathbf{D}$ is symmetric, these terms determine the remaining off-diagonal entries of the tensor. Mixed derivatives incorporate multiple coordinate directions, and their discretization produces sign-indefinite (positive or negative) stencil contributions. Therefore, these terms are not grouped with the positivity-preserving diagonal diffusion terms. These are instead discretized with centered differences and added through the correction step described below. The drift like term, $\nabla \cdot(\phi (\nabla^T\mathbf{D})C_j)$, behaves like an advection term. We discretize it using an upwind approximation at cell faces for stability. However, because the stencil is not guaranteed to have the same sign structure as the face-averaged diagonal diffusion operator, it is also added later through a correction step rather than grouped with positivity-preserving diagonal diffusion terms to form the backward Euler matrix.

The internalization term is linear and contributes only to the diagonal of the discrete operator and strengthens the M-matrix structure of the backward Euler solve. For this reason, the NSC internalization term is grouped with the positivity-preserving part of the method. The production term is handled according to its dependence on $q_j$. Two types of production forms are considered. In the Gaussian initial condition case with the tumor-sourced production term, there is a part that reinforces the diagonal dominance of the M-matrix that involves some negative multiple of $q_j$ and another part that is additive and without $q_j$, which is treated as a forcing term on the right-hand side of the matrix equation. In the constant initial condition case, which has a logistic-type production term, the whole term does not contribute to the M-matrix structure and has a nonlinear part, so it is treated entirely explicitly using the solution at the prior time step and added to the forcing term.

The discrete operator, $L$, representing all terms in the discretized PDE except the forcing terms, can be split as $L = L_{\text{low}}+ L_{\text{correction}}$. $L_{\text{low}}$ is the operator containing all of the terms with the M-matrix structure that preserve nonnegativity of $C_j$. These terms include the finite-volume diagonal diffusion operator, the NSC internalization term, and part of the production term in the PDE under the Gaussian initial condition case. $L_{\text{correction}}$ includes the other terms not in $L_{\text{low}}$ or treated as a forcing term, which include the centered difference cross-diffusion terms and the upwinded drift term. The forcing term includes part or all of the production term, depending on the type of production.

For computational efficiency on such a large domain, the resulting sparse linear system from the low-order implicit-explicit method is solved using BiCGSTAB (biconjugate gradient stabilized method) with an incomplete LU preconditioner. After the positivity-preserving low-order solve, the remaining terms are added to the low-order solution through a global positivity-preserving correction step. We add in the correction term $\Delta tL_{\text{correction}} q_{\text{low}}^{n-1}$ scaled by a global scalar $\theta \in [0,1]$, chosen so that no component of $q_j^{n+1}$ becomes negative. }


\section*{Acknowledgments}
\blue{
The authors thank the anonymous reviewers for their valuable suggestions. 
}
\blue{
\section*{Data availability} 

\noindent 
Raw data of the mouse brain DTI were obtained by Chuang Lab Atlas repository \cite{chuang2011mri}. 
The code will be available in https://github.com/apharri7/NSC-Migration-abm-pde upon publication. 
}


\bibliographystyle{unsrt}
\bibliography{NSC.bib}

\end{document}